\documentclass[dvips]{elsart}
\usepackage{graphicx}    

\begin{document}


\begin{frontmatter}

%
%
\title{ EMMA -- A New Underground Cosmic-Ray Experiment }

%
%
\author[cuppp]{T.~Enqvist\corauthref{tc}}, \ead{timo.enqvist@oulu.fi}
\author[cuppp]{V.~F\"ohr},
\author[cuppo]{J.~Joutsenvaara},
\author[sgo]{T.~J\"ams\'en},
\author[jyfl]{P.~Ker\"anen},
\author[cuppp]{P.~Kuusiniemi},
\author[cuppo]{H.~Laitala},
\author[cuppp]{M.~Lehtola},
\author[cuppo]{A.~Mattila},
\author[cuppp]{J.~Narkilahti},
\author[cuppo]{J.~Peltoniemi},
\author[cuppo]{H.~Remes},
\author[cuppp]{M.~Reponen},
\author[cuppo]{T.~R\"aih\"a},
\author[cuppo]{J.~Sarkamo},
\author[cuppp]{C.~Shen},
\author[sgo]{I.~Usoskin},
\author[cuppo]{M.~Vaittinen},
\author[cuppp]{Z.~Zhang},
\author[ihep]{L.~Ding}, 
\author[ihep]{Q.~Zhu}, 
\author[hyfl]{M.~Roos}, 
\author[inr]{I.~Dzaparova}, 
\author[inr]{S.~Karpov}, 
\author[inr]{A.~Kurenya}, 
\author[inr]{V.~Petkov}, 
\author[inr]{A.~Yanin},
\author[arhus]{H.~Fynbo}
%
%
\address[cuppp]{Centre for Underground Physics (CUPP), 
P.O. Box 22, FIN-86801 Pyh\"asalmi, Finland}
\address[cuppo]{CUPP, P.O. Box 3000, FIN-90014 University of Oulu, Finland}
\address[sgo]{Sodankyl\"a Geophysical Observatory (SGO), University of Oulu, 
Finland}
\address[jyfl]{Department of Physics, University of Jyv\"askyl\"a, Finland}
\address[ihep]{IHEP, Chinese Academy of Sciences, Beijing, China}
\address[hyfl]{Department of Physical Sciences, University of Helsinki, 
Finland}
\address[inr]{INR, Russian Academy of Sciences, Moscow, Russia}
\address[arhus]{Department of Physics and Astronomy, University of Aarhus, 
Denmark}
\corauth[tc]{Corresponding author.}



\begin{abstract}

A new type of cosmic-ray experiment is under construction in the Pyh\"asalmi 
mine in the underground laboratory of the University of Oulu, Finland. It 
aims to study the composition of cosmic rays at and above the \textit{knee} 
region. The experiment, called EMMA, will cover approximately 150 m$^2$ of 
detector area. The array is capable of measuring the multiplicity and the 
lateral distribution of underground muons, and the arrival direction of the 
air shower. The full-size detector is expected to run by the end of 2007.

\end{abstract}

\end{frontmatter}

\section{Introduction}  

Due to a slight change observed in the cosmic-ray energy spectrum in 
the energy interval of 10$^{15}$ -- 10$^{16}$ eV, it is believed that 
the origin, modification in the chemical composition, acceleration 
mechanism or propagation of cosmic rays (or a combination of these) 
changes. Up to this energy, so called \textit{knee} region, most cosmic 
rays are supposed to be produced inside the galaxy, and are also confined 
by the galactic magnetic field.

At these high energies the source cannot be observed directly and the 
cosmic-ray composition is used as a tool to investigate the origin of 
the cosmic radiation. The direct composition measurements are no longer 
practical at or above the knee and the method is solely based on the 
measurement of extensive air showers, i.e. the secondary particles 
created in the atmosphere and detected by large arrays on the ground. 

The origin of the knee has been one of the fundamental problems of 
cosmic-ray physics, and it has been discussed for decades. Several 
models have been presented predicting different composition at the knee 
energies, and could only be identified by the experimental evidence on 
the composition. Some new experimental efforts have been devoted to the 
study of cosmic rays in recent years. These experiments are based, for 
example, on multi-parameter measurement of extensive air-showers, on 
shower maximum measurement by \v Cerenkov or fluorescence detectors and 
on underground multimuon measurements (see, for example, Ref. \cite{Hor05} 
and references therein). Their conclusions, however, have so far been 
diverse, implying the need for further studies, especially using different 
approaches. The results are also known to be strongly model dependent.

EMMA (Experiment with MultiMuon Array) uses a different approach. It is 
not the first underground cosmic-ray experiment (see, for example, Refs. 
\cite{Ceb90,Kas97,Ava03,Gru03,EAS04}), but it differs significantly from 
previous underground experiments with its ability to measure the lateral 
distribution function of underground muons. In EMMA the composition 
analysis is based on the lateral distribution of high-energy muons and 
on their multiplicity. The muons detected by EMMA are generated in the
upper part of the air shower close to the primary interaction.

\section{Experimental details}  

The present experiment will be carried out at the depth of 85 metres in
the Pyh\"asalmi mine (owned by the Inmet Mining Corporation Ltd., Canada),
which is situated in the middle of Finland.

\begin{figure}
 \begin{center}
  \leavevmode
  \fbox{ \includegraphics[scale=0.55, angle=-90]{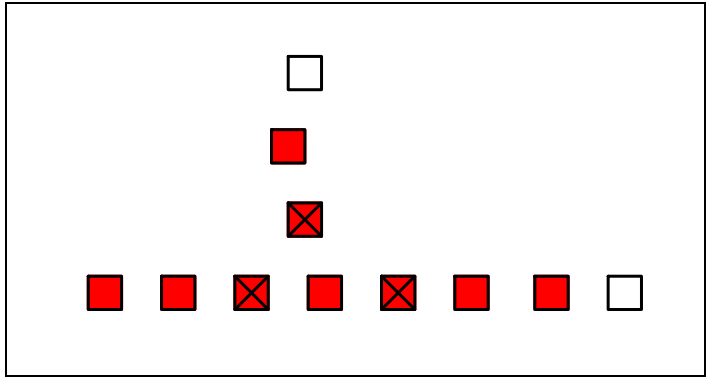} }
  \caption{
   Schematic layout of the EMMA array in the underground site at 85 metres. 
   The array consists of nine units (red squares) each having an area of
   about 15 m$^2$. The horizontal and vertical dimensions are approximately 
   50 and 20 metres, respectively. The three crossed units occupy detectors 
   in two layers for shower direction measurement. The gaps between the
   units can be filled with detectors and the two open squares indicate
   the possible additional units if the array is expanded. 
  }
  \label{figu_layout} 
 \end{center}
\end{figure}

The detector array is placed at the depth of 85 metres (corresponding 
240 m.w.e) which gives a threshold energy of muons of approximately 50 GeV.
The rock overburden filters out all other particles of the air shower except
the high-energy muons (and neutrinos). A schematic layout of the array is 
shown in Fig. \ref{figu_layout}, where each of the nine units have an area 
of about 15 m$^2$. The area between the units is close to the unit area.  

The EMMA detector array consists of drift chambers (muon barrel (MUB)
detectors) previously used in the LEP--DELPHI experiments at CERN 
\cite{Delphi}. The setup is able to measure the arrival direction of the 
air shower, the muon multiplicity and their lateral distribution. This is 
carried out by an array of separate detectors with the total detector area 
of about 150 m$^2$. Part of the array is in two layers with a vertical 
distance of 2.5 metres between the layers in order to obtain the direction 
information (see Fig. \ref{figu_layout} for details).

Most of the drift chambers have an active volume of 365$\times$20$\times$1.6 
cm$^3$.  A plank consists of seven chambers (partly overlapped) having an 
area of approximately 3 m$^2$. The layout of Fig. \ref{figu_layout} requires 
60 planks. The drift chambers operate in the proportional mode, with 
Ar:CO$_{2}$ (92:8) nonflammable gas mixture. Due to the safety issues, the 
gas mixture does not contain CH$_4$. Each drift chamber can provide up to 
three signals, one anode signal and two delay line signals (near and far), 
which can be used to localise the positions of particles passing through 
the chambers. The measured position resolution of the chambers in our setup 
is about 1 cm and 3 cm along the drift direction and along the drift line,
respectively.

Recycling the old muon barrel detectors provides a the possibility to built 
the array at low costs. Also the use of existing caverns of the mine reduces 
the costs. The data-acquisition electronics is build by ourselves or bought 
as new.

The underground array is under construction and the data recording can be
started around middle of 2006 with a partial-size array. The full-size array 
is expected to be in operation by the end of 2007.

\section{Data analysis}  

\subsection{Muon tracking}  

An important part of the data analysis software is the muon tracking
routine. As it is obvious that an observed shape of the shower partially
depends on the angles it hits the detectors, the shower can only be
reconstructed by determining the angles of individual muons. In EMMA the
shower reconstruction will be carried out by tracking muons as they hit 
simultaneously to two detector planes (the three double-layer units shown
in Fig. \ref{figu_layout}).

\begin{figure}
 \begin{center}
  \leavevmode
  \includegraphics[scale=0.62,angle=-90]{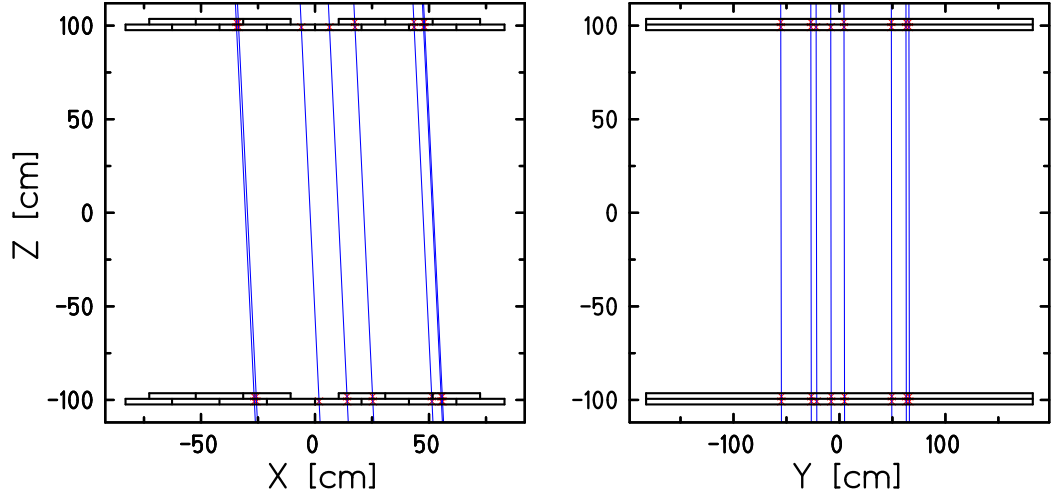}
  \caption{
   Eight muons hitting two planes of detectors with randomly chosen shower 
   direction and muon hit positions. The lines connecting the hits in the 
   upper and lower planes result from the tracking code. 
  }
  \label{figu_track}  
 \end{center}
\end{figure}

The simulated performance of the muon tracking routine is illustrated in 
Fig.~\ref{figu_track} using an example where eight simultaneous muons 
penetrate two detector planes with a randomly chosen shower angle. The 
tracking code employs an automatic routine which connects the pairs in 
two planes and rejects pairs too far from the average angle. 

The tracking routine will be tested with real cosmic-ray muon data at the
surface in the end of 2005 and in the beginning of 2006 using system similar 
to Fig.~\ref{figu_track}, and it will be further developed.

\subsection{Locating core position}  

\begin{figure}
 \begin{center}
  \leavevmode
  \includegraphics[scale=0.8,angle=-90]{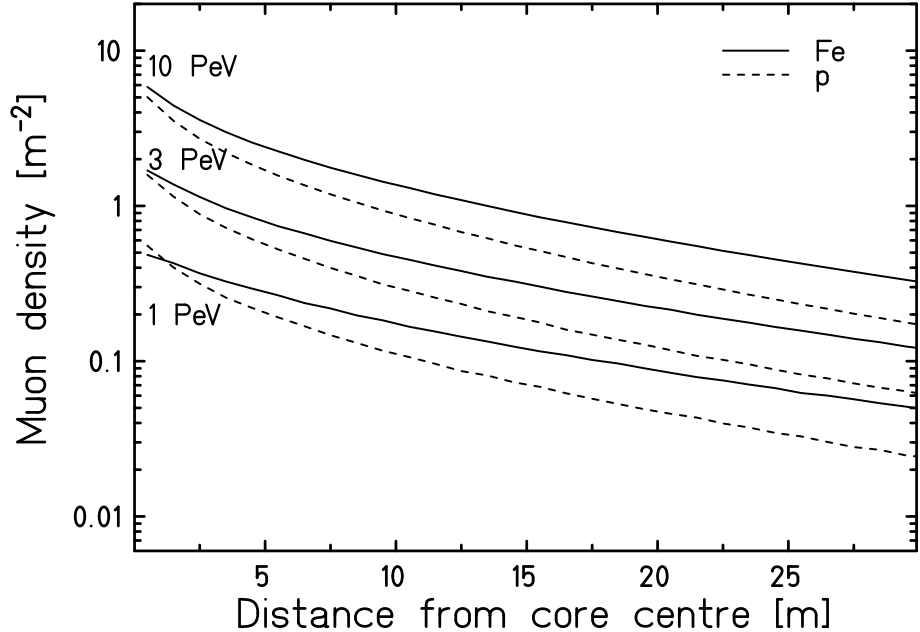}
  \caption{
   Simulated (CORSIKA \& QGSJET \cite{Hec98,Kal97}) lateral distributions 
   of muons for 1, 3 and 10 PeV proton- and iron-initiated showers. The 
   cutoff in muon energy is 50 GeV.
  }
  \label{figu_ldfs}  
 \end{center}
\end{figure}

In order to reconstruct the lateral distribution of underground muons 
and their multiplicity, the core of the air shower has to be located 
accurately (i.e. within metres). A two-dimensional fit routine was 
developed for the extraction of the core location. The routine was tested
with simulated air-shower samples. Depending on the hit position of the 
core, it can be extracted with an accuracy better than three metres in 
most cases if the core hits the detector unit or between them. If the core 
is elsewere, the accuracy is weaker and the distribution slightly biased, 
but this can be corrected for. In the current version of the core-location 
software the routine uses model-dependent lateral distributions. However, 
we are developing a routine that works in the model-independent way, which
would allow a direct comparison between measured lateral distribution and 
different model predictions.

\subsection{Extracting composition}

The multiplicity measurements at distances between 20 and 30 metres from 
the core position are expected to be the optimal cases for separating 
different cosmic-ray primary particles. As the core is located and the 
lateral distribution is fitted for each shower, the measured lateral 
distributions are used to extract the composition information as a 
function of the primary energy. The number of muons at the core can be 
used as an indicator for the primary energy. The energy resolution of EMMA 
according to this preliminary analysis is somewhat moderate.

\section{Expected results}  

As no experimental results have been obtained yet, expected results based 
on simulations are introduced in this section to present some of the 
performance of the EMMA array. The simulations serve, at the same time, 
also a way to develop and further improve the methods to analyse and to 
interprete the data. The air-shower simulations and particularly the data 
analysis methods are still preliminary but they already indicate the good
capabilities of the EMMA array.

\begin{figure}
 \begin{center}
  \leavevmode
  \includegraphics[scale=0.75,angle=-90]{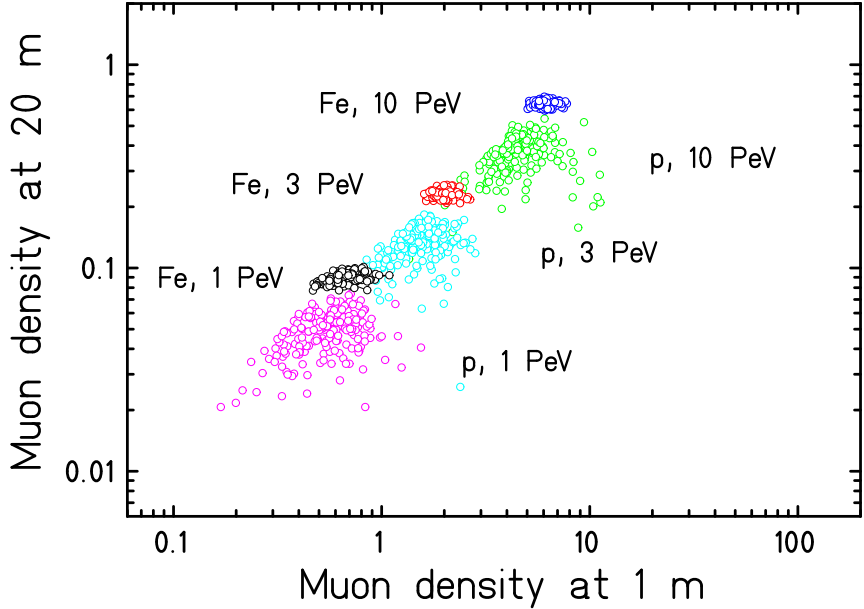}
  \caption{
   The separation of primary cosmic-ray composition and energy, in 
   two-dimensional plot of $\rho$(1 m) versus $\rho$(20 m). See text for
   details. The plot is based on simulations of proton and iron primaries 
   with energies of 1, 3 and 10 PeV.
  }
  \label{figu_showers}  
 \end{center}
\end{figure}

According to cosmic-ray air-shower models (e.g. CORSIKA \& QGSJET 
\cite{Hec98,Kal97}) the lateral distribution, or its gradient, of 
high-energy muons is sensitive to primary cosmic-ray particles and their 
energies, and to hadronic interactions. Also the muon multiplicity is 
sensitive to the energy. The simulated (using CORSIKA \& QGSJET) lateral 
distribution functions of muons are shown in Fig. \ref{figu_ldfs} for three 
discrete energies of 1, 3 and 10 PeV for proton- and iron-initiated showers,
being the most relevant for the present experiment. Only muons with energies
above 50 GeV are included in Fig. \ref{figu_ldfs}.

The method of the cosmic-ray composition study in the present work is 
illustrated in Fig. \ref{figu_showers} where the separation of two primary 
cosmic-ray particles with three different energies is shown. The figure is 
generated by fitting the lateral distribution functions to muons in an air 
shower and extracting the muon densities at distances of 1 and 20 metres 
($\rho$(1 m) and  $\rho$(20 m), respectively) away from the located core 
position. The composition and energy of the primary cosmic ray can then be 
extracted, for example, using variables $\rho$(1 m), which is related to 
the energy, and $\rho$(20 m), which is related to the composition.

In reality, different groups (or groups of elements) in the plot would be 
less pronounced than those simulated and shown in Fig. \ref{figu_showers}. 
This is due to poorer statistics and an uncertainty related to the core 
position determination, and also continuous cosmic-ray primary energy 
spectrum (Fig. \ref{figu_showers} shows only three discrete energies for
only two primary particles). 

\begin{figure}
 \begin{center}
  \includegraphics[scale=0.75,angle=-90]{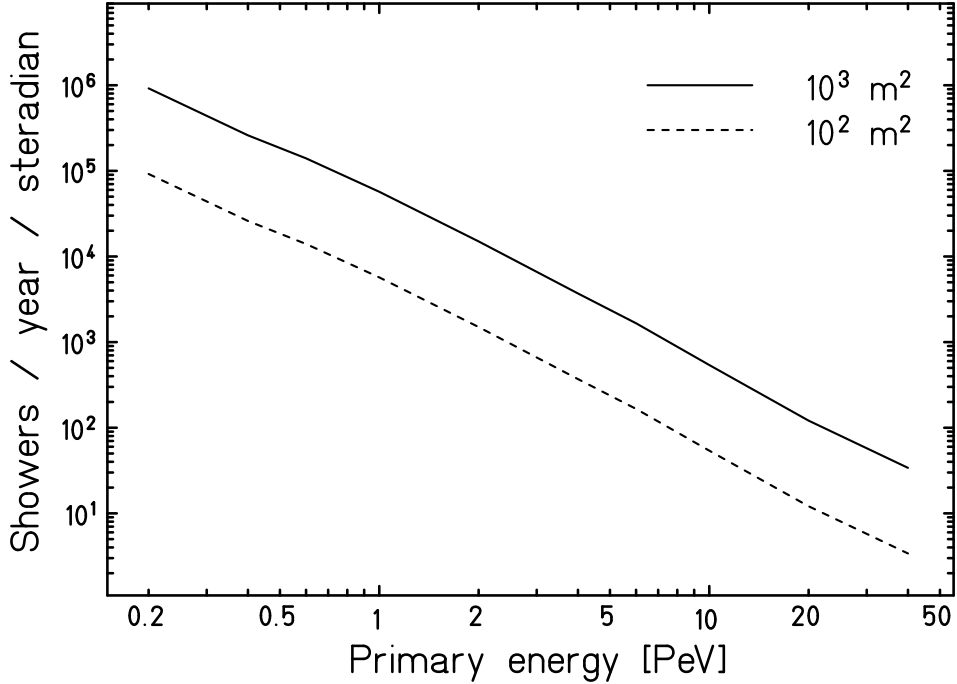}
  \caption{
   Integrated shower rates above given primary energies for two effective 
   detection areas.
  }
  \label{figu_stats}  
 \end{center}
\end{figure}

The effective area of the array determines recorded statistics. The area in
the EMMA experiment is expected to be between 100 m$^2$ and 1000 m$^2$. Fig. 
\ref{figu_stats} shows the expected numbers of air showers collected as a
function of cosmic-ray primary energy. 

\section{Summary}

A new underground cosmic ray experiment EMMA is under construction and it
is expected to start recording data in the full scale by the end of 2007. 
With a partial-size array the data recording can be started by the middle 
of 2006. The analysis of simulated air showers shows that the primary
cosmic-ray composition could be resolved (with a two-component model) at
and above the knee energies. A possibility for the model-independent way of 
the determination of the muon lateral distribution would allow to improve
high-energy interaction models. Due to new method used, the EMMA experiment 
could provide comprehensive (and perhaps new) information on the composition 
of the cosmic rays at the knee region within the next few years.

\ack

The support from the Magnus Ehrnrooth Foundation, the Jenny and Antti Wihuri 
Foundation, the Finnish Academy of Science and Letters (V\"ais\"al\"a 
Foundation), and the Finnish Cultural Foundation is acknowledged.
The work is funded by the European Union Regional Development Fund and it 
is also supported by the Academy of Finland (projects 108991, 7108875 and
7106570).


\begin{thebibliography}{abc00}

\bibitem{Hor05} 
 J\"org R. H\"orandel, 
 astro-ph/0508014 (2005).
\bibitem{Ceb90} 
 D. Cebula, S.C. Corbat\'o, T. Daily, D.B. Kieda, K. Lande, C.K. Lee,
 M.L. Cherry,
 The Astrophysical Journal 358 (1990) 637.
\bibitem{Kas97} 
 S.M. Kasahara \textit{et al.},
 Physical Review D55 (1997) 5282.
\bibitem{Ava03} 
 V. Avati, L. Dick, K. Eggert, J. Str\"om, H. Wachsmuth, S. Schmeling,
 T. Ziegler, A. Br\"uhl, C. Grupen, 
 Astroparticle Physics 19 (2003) 513.
\bibitem{Gru03} 
 C. Grupen, M.T. Kurt, A. Mailov, A.S. M\"uller, A. Putzer, B. Rensch,
 H.G. Sander, S. Schmeling, M. Schmelling, H. Wachsmuth, T. Ziegler,
 K. Zuber,
 Nuclear Instruments and Methods A510 (2003) 190.
\bibitem{EAS04} 
 EAS-TOP Collaboration \& MACRO Collaboration (M. Aglietta \textit{et al.}),
 Astroparticle Physics 20 (2004) 641.
\bibitem{Delphi} 
 DELPHI Collaboration,
 Nuclear Instruments and Methods A303~(1991)~233.
\bibitem{Hec98} 
 D. Heck \textit{et al.}, Report Forschungzentrum Karlsruhe 6019 (1998).
\bibitem{Kal97} 
 N.N. Kalmykov \textit{et al.},
 Nuclear Physics B (Proc. Suppl.) 52B (1997) 17.
\end{thebibliography}
\end{document}